\documentclass[aps,prd,preprint,groupedaddress,nofootinbib,amsmath,showkeys,tightenlines]{revtex4}

\newcommand{\ul}[1]{\hspace{-0.1ex}\underline{\hspace{0.1ex} #1 \hspace{-0.1em}}\hspace{0.1em}}
\newcommand{\tauindicesstring}{underlining }

\newcommand{\bydef}{\equiv} 

\newcommand{\ts}{\textstyle}
\newcommand{\ds}{\displaystyle}
\newcommand{\be}{\begin{equation}}
\newcommand{\ee}{\end{equation}}
\newcommand{\bea}{\begin{eqnarray}}
\newcommand{\eea}{\end{eqnarray}}
\newcommand{\bse}{\begin{subequations}}
\newcommand{\ese}{\end{subequations}}
\newcommand{\ifhat}{} 

\newcommand{\pp}{\pi} 
\newcommand{\kk}{\kappa} 
\newcommand{\hpp}[1]{{\ifhat \pp_{#1}}} 
\newcommand{\hkk}[1]{{\ifhat \kk^{#1}}} 
\newcommand{\opppp}[1]{{\widehat{\left(#1\right)}_{\pp\pp}}} 
\newcommand{\okkkk}[1]{{\widehat{\left(#1\right)}_{\kk\kk}}} 
\newcommand{\oppkk}[1]{{\widehat{\left(#1\right)}_{\pp\kk}}} 
\newcommand{\smopppp}[1]{{\widehat{(#1)}_{\pp\pp}}} 
\newcommand{\smokkkk}[1]{{\widehat{(#1)}_{\kk\kk}}}
\newcommand{\smoppkk}[1]{{\widehat{(#1)}_{\pp\kk}}}
\newcommand{\fto}[1]{{\hat a_{#1}}} 
\newcommand{\ftoa}[1]{{\hat a^\dagger_{#1}}} 

\newcommand{\quadalg}{{\cal A}_2}

\newcommand{\DSx}{Y} 
\newcommand{\DS}[1]{Y_{\ul #1}} 

\usepackage{bm}

\begin{document}

\title{Extension of \textit{\textbf{C}}(1,3) (Super)Conformal
Symmetry using Heisenberg and Parabose operators}

\author{Igor Salom}
\email{isalom@phy.bg.ac.yu}
\affiliation{Institute of Physics,
11001 Belgrade, P.O. Box 57, Serbia}

\date{\today}

\begin{abstract}
In this paper we investigate a particular possibility to extend
$C(1,3)$ conformal symmetry using Heisenberg operators, and a
related possibility to extend conformal supersymmetry using
parabose operators. The symmetry proposed is of a simple
mathematical form, as is the form of  necessary symmetry breaking
that reduces it to the conformal (super)symmetry. It turns out
that this extension of conformal superalgebra can be obtained from
standard non-extended conformal superalgebra by allowing
anticommutators $\{Q_\eta, Q_\xi\}$ and $\{\overline Q_{\dot
\eta}, \overline Q_{\dot \xi}\}$ to be nonzero operators and then
by closing the algebra. In regard of the famous Coleman and
Mandula theorem (and related Haag-Lopuszanski-Sohnius theorem),
the higher symmetries that we consider do not satisfy the
requirement for finite number of particles with masses below any
given constant. However, we argue that in the context of theories
with broken symmetries, this constraint may be unnecessarily
strong.
\end{abstract}


\maketitle

\section{\label{sec1}Introduction}

Prospect of finding larger symmetries that would embed observable
Poincare symmetry and possibly some of the internal symmetries (in
a non trivial way) has attracted physicists for a long time. Early
attempts ended in formulation of famous Coleman and Mandula no-go
theorem \cite{CM}, but only to be soon evaded by the idea of
supersymmetry. The Coleman and Mandula theorem was then replaced
by Haag-Lopuszanski-Sohnius theorem \cite{HLS} that put the now
standard super-Poincare and super-conformal symmetries at the
place of maximal supersymmetries of realistic models (up to
multiplication by an internal symmetry group). However, the
attempts to go around these no-go theorems never truly ceased,
mostly trying to weaken the mathematical requirements of the
theorems \cite{ZaobilazenjaHLS1, ZaobilazenjaHLS2}.

We will here consider an extension of conformal (super)symmetry
which does not meet one of the physical premises of the theorem
\cite{CM} - namely the "particle finiteness" premise: "for any
finite M, there are only finite number of particle types with mass
less than M". Regarding this requirement, S. Coleman in paper
\cite{Cpretecha} comments: "We would probably be willing to accept
a theory with an infinite number of particles, as long as they
were spread out in mass in such a way that experiments conducted
at limited energy could only detect a finite number of them". Our
point is that a proper symmetry breaking can, in principle, induce
such mass splitting. Besides, this does not have to imply
increasing of the complexity of a theory, since symmetry breaking
is already an inescapable component of most of the contemporary
physical models.

The mathematical motivation for choice of the particular
symmetries that we are going to investigate comes from the fact
that $C(1,3)$ conformal symmetry is in an interesting way
contained in algebra that is formed by all (hermitian) quadratic
polynomials of four pairs of Heisenberg operators [this algebra,
constructed in the following section, is isomorphic to $sp(2n)$
where $n=4$]. Namely, reduction from $sp(8)$ to $c(1,3)$ can be
seen as a consequence of one $SU(2)$ to $U(1)$ symmetry breaking.
Though this construction based on Heisenberg operators has many
interesting properties itself, we will actually use it as an
intermediate step, serving to introduce bosonic part of a larger
symmetry.

In the next step (section \ref{sec3}) we will show that by
replacing the starting Heisenberg algebra with relations of
parabose algebra, we arrive to an extension of super-conformal
algebra. Relation of this algebra with standard conformal
superalgebra is rather interesting: it is the algebra that is
obtained from conformal superalgebra when we remove the algebraic
"constraints" $\{Q_\eta, Q_\xi\} = 0$ (allowing these and adjoint
anticommutators to be new symmetry generators) and appropriately
close the algebra. Intriguing is that enlarging the conformal
superalgebra in this way simplifies the algebra instead of
complicating it, as the structural relations of the larger
symmetry are determined by two defining relations of parabose
algebra.

As consecutive action of the same supersymmetry generator no
longer annihilates a state, it is obvious that number of particles
in a supermultiplet becomes infinite. However, we argue that such
an obvious disagreement with experimental data is a problem of the
qualitatively same type as occurs in the standard Poincare
supersymmetry. Namely, whereas in models with Poincare
supersymmetry we need a symmetry breaking to induce mass
differences between finite number of super-partners, here the
symmetry breaking should provide ascending masses for infinite
series of super-partners. We will demonstrate existence of simple
form of symmetry breaking that reduces the symmetry down to
Poincare group, altogether with providing the mass splitting.

In this paper we do not intend to present details of any realistic
model, but rather to point out some theoretical possibilities
which are, to our opinion, unrightfully ignored. In the moment
when experimental confirmation of existence of supersymmetry is
being expected, we should be aware of all possible variations of
supersymmetry idea.

Throughout the text, Latin indices $i, j, k, \dots$ will take
values 1, 2 and 3, Greek indices from the beginning of alphabet
$\alpha, \beta, \dots$ will take values from 1 to 4 and will in
general denote Dirac-like spinor indices, $\eta$ and $\xi$ will be
two-dimensional Weyl spinor indices, while Greek indices from the
middle of alphabet $\mu, \nu, \dots$ will denote Lorentz
four-vector indices.

\section{\label{sec2}Heisenberg Operators and Extension of Conformal Symmetry}

Let operators $\hkk{\alpha}$ and $\hpp{\alpha}$ satisfy Heisenberg
algebra in four dimensions\footnote{In spite of this, we stress
that these operators do not represent coordinates and momenta.
Furthermore, they will turn out to transform like Dirac spinors.}: %
\be [\hkk{\alpha}, \hpp{\beta}] = i \delta_\beta^\alpha,
[\hkk{\alpha}, \hkk{\beta}] = [\hpp{\alpha}, \hpp{\beta}] = 0.\ee
There are three types of quadratic combinations of these
operators: quadratic in $\hkk{\alpha}$, quadratic in
$\hpp{\alpha}$ and mixed. Hermitian operators of each of these
kinds can be written in matrix notation, respectively as: %
\bea \okkkk{A} &\bydef& A_{\alpha\beta} {\textstyle\frac12}\{\hkk{\alpha},\hkk{\beta}\}, \nonumber \\%
\opppp{A} &\bydef& A^{\alpha\beta}
{\textstyle\frac12}\{\hpp{\alpha},\hpp{\beta}\},
\label{quadratic operators} \\
\oppkk{A} &\bydef& A^\alpha_{\ \beta}
{\textstyle\frac12}\{\hpp{\alpha},\hkk{\beta}\}, \nonumber
\eea %
where $A$ is an arbitrary four by four real matrix, with
restriction that matrices appearing in definitions $\okkkk{A}$ and
$\opppp{A}$ are implied to be symmetric.\footnote{A hat sign over
a matrix will be used to emphasize the difference between the
operator obtained from a matrix in the sense of definition
(\ref{quadratic operators}) and the matrix itself.}

Such quadratic operators form an algebra with commutation
relations easily derivable from the Heisenberg algebra relations.
This \emph{algebra of quadratic operators} $\quadalg$ we wish to
consider as an extension of conformal algebra.

To reveal the conformal subalgebra in this structure, first we
choose a set of six real matrices $\sigma_i$ and $\tau_{\ul i}$,
$i, \ul i = 1,2,3$ (four dimensional analogs of Pauli matrices) satisfying %
\be [\sigma_i, \sigma_j] = 2\:\! \varepsilon_{ijk} \sigma_k,\quad %
[\tau_{\ul i}, \tau_{\ul j}] = 2 \:\! \varepsilon_{\ul i \ul j \ul k} \tau_{\ul k},\quad %
[\sigma_i, \tau_{\ul j}] = 0, %
\label{sigma tau commutators} \ee %
as a basis of antisymmetric four by four real
matrices\footnote{One possible realization of such matrices is,
for example: $\sigma_1 = -i\sigma_y \times \sigma_x$, $\sigma_2 =
-i I_2 \times \sigma_y$, $\sigma_3 = -i\sigma_y \times \sigma_z$,
$\tau_1 = i\sigma_x \times \sigma_y$, $\tau_2 = -i \sigma_z \times
\sigma_y$, $\tau_3 = -i\sigma_y \times I_2$, where $\sigma_x$,
$\sigma_y$ and $\sigma_z$ are standard two dimensional Pauli
matrices and $I_2$ is a two dimensional unit matrix.} (we distinct
tau indices from sigma indices by \tauindicesstring the former).
However, unlike Pauli matrices, these matrices are anti-hermitian,
satisfying $\sigma_i^2 = \tau_{\ul i}^2 = -1$. As a basis for
symmetric matrices we choose nine matrices $\alpha_{\ul ij} \bydef
\tau_{\ul i} \sigma_j$ and unit matrix denoted as $\alpha_0$.

Now, set of 36 operators %
\be \Big\{\smoppkk{\tau_{\ul i}}, \oppkk{\sigma_j},
\oppkk{\alpha_0}, \smoppkk{\alpha_{\ul ij}},
 \opppp{\alpha_0}, \smopppp{\alpha_{\ul ij}}, \okkkk{\alpha_0},
\smokkkk{\alpha_{\ul
ij}}\Big\} \label{Jeleb-conformal basis} \ee %
can be chosen as basis of algebra of quadratic operators.

To obtain conformal subalgebra let us discard all operators from
this set with underlined index having values $\ul 1$ and $\ul 2$.
What we are left with is a subalgebra isomorphic with conformal
algebra $c(1,3)$ plus one additional generator that commutes with
the rest of the subalgebra. Thus we introduce new notation for the
remaining generators: %
\bea & & \ifhat J_k \bydef \oppkk{\displaystyle
\frac{\sigma_i}{2}}, \quad \ifhat N_i \bydef
\oppkk{\displaystyle \frac{\alpha_{\ul 3 i}}{2}}, \quad  %
\ifhat D \bydef \oppkk{\displaystyle \frac{\alpha_0}{2}},
  \nonumber \\ %
& & \ifhat P_i \bydef \opppp{\displaystyle \frac{\alpha_{\ul 3
i}}{2}}, \quad \ifhat P_0 \bydef \opppp{\displaystyle \frac{
\alpha_0 }{2}}, \nonumber \\ %
& & \ifhat K_i \bydef \okkkk{\displaystyle \frac{\alpha_{\ul 3
i}}{2}}, \quad \ifhat K_0 \bydef -\okkkk{\displaystyle
\frac{\alpha_0}{2}}, \label{conformal identification} \eea %
where $J_i, N_i, D, P_\mu$ and $K_\mu$ play roles of rotation
generators, boost generators, dilatation generator, momenta and
pure conformal generators, respectively. The additional remaining operator is %
\be \ifhat \DS{3} \bydef \oppkk{\ds \frac{\tau_{\ul 3}}{2}},
\label{helicity generator}\ee%
which commutes with all of the conformal generators. We will call
it the third component of the \emph{dual spin}.

If we consider the way in which the recognized conformal
subalgebra fits into the larger algebra $\quadalg$, we see that
spatial momenta, being equal to $\smopppp{ \frac{\alpha_{\ul 3
j}}{2}}$ naturally fit into a set of nine operators $\smopppp{
\frac{\alpha_{\ul ij}}{2}}$, spatial components of pure conformal
generators fit into a set of nine $\smokkkk{ \frac{\alpha_{\ul
ij}}{2}}$, boosts into set of nine $\smoppkk{ \frac{\alpha_{\ul
ij}}{2}}$ and the third component of the dual spin fits into set
of three operators $\smoppkk{ \frac{\tau_{\ul i}}{2}}$. Having
this on mind, we can extend notation
convention (\ref{conformal identification}) to cover whole algebra $\quadalg$: %
\be \ifhat N_{\ul ij} \bydef \smoppkk{\displaystyle
\frac{\alpha_{\ul ij}}{2}},\ifhat P_{\ul ij} \bydef
\smopppp{\displaystyle \frac{\alpha_{\ul ij}}{2}},\ifhat K_{\ul
ij} \bydef \smokkkk{\displaystyle \frac{\alpha_{\ul ij}}{2}},
\ifhat \DS{i} \bydef \smoppkk{\ds \frac{\tau_{\ul i}}{2}},
\label{Jeleb identification} \ee %
while for spatial components of conformal generators it holds $N_i
= N_{\ul 3i}$, $P_i = P_{\ul 3i}$ and $K_i = K_{\ul 3i}$.

Alternatively, we could have obtained conformal subalgebra by
keeping operators with underlined index equal to 1 or 2, instead
3. As the matter in fact, if we pick any linear combination of
operators $\DS{i}$, or of operators $J_i$, the subalgebra of
$\quadalg$ that commutes with the chosen operator will be $c(1,3)
$ isomorphic. On the other hand, operators $\ifhat \DS{i}$ and
$\ifhat J_i$ constitute two, mutually commuting $su(2)$ isomorphic
subalgebras (a consequence of $so(4) = su(2) \oplus su(2)$
identity). The two corresponding $SU(2)$ isomorphic groups act,
respectively, on underlined and on non-underlined indices of
algebra operators. The situation slightly looks like as if we had
two independent rotation groups, while "momenta", "boosts" and
"pure conformal operators" were here determined by two independent
three-vector directions, each related to its own "rotation" group.
And the symmetry reduction from $\quadalg$ group to its
conformally isomorphic subgroup can be therefore understood as a
consequence of symmetry breaking of one of these two $SU(2)$
subgroups. Without loss of generality, we have assumed breaking of
the group generated by $\ifhat \DS{i}$, with $\ifhat \DS{3}$
generating the remaining $U(1)$ symmetry.

As a more concrete example of such symmetry breaking, we can
assume existence of effective potential being an increasing
function of absolute value of $\DS{3}$ [e.g.\ proportional to the
$(\DS{3})^2$]. If the potential is sufficiently strong, all low
energy physics would be constrained to subspace of $\DS{3}$
eigenvalue equal to zero, and the remaining symmetry would be
conformal symmetry. Moreover, since such a potential would have to
break dilatational symmetry, overall symmetry would be reduced to
the observable Poincare group. (Unfortunately, this particular
symmetry breaking example seems to be oversimplified for more
delicate physical reasons that we will not analyze in this paper.)

Note that such symmetry breaking also automatically fixes metric
of space-time (i.e.\ of the remained symmetry) to be Minkowskian
[the remaining $J_i$ and $N_{\ul 3i}$ hermitian operators form
exactly $so(1,3)$ algebra]. It is also interesting that energy
operator $P_0$ singles out among other momentum operators (i.e.\
among the rest of operators quadratic in $\hpp{}$) even before the
symmetry reduction. Indeed, this operator, being the sum of
squares of $\hpp{\alpha}$, stands out as a positive operator, and
there is no algebra automorphism that takes any other "momentum
component" $P_{\ul ij}$ into the $P_0$ or vice-versus. This gives
us some right to interpret the full group generated by $\quadalg$
as a symmetry that differs from the observable space-time symmetry
in the first place by existence of two "spatial-like" rotations,
whereas it possesses something that looks like unique role of one
axis (to be interpreted as the time axis). And the symmetry
breaking only gets us rid of one of the "rotation-like" groups.

Structural relations of algebra $\quadalg$ are:%
\bea && [\ifhat J_i, \ifhat J_j] = i\, \varepsilon_{ijk} \ifhat
J_k,\quad [\ifhat \DSx_{\ul i}, \ifhat \DSx_{\ul j}] =  i\,
\varepsilon_{\ul i\ul j\ul k} \ifhat \DSx_{\ul k},
\quad [\ifhat J_i, \ifhat \DSx_{\ul j}] = 0, \nonumber \\
{}&& [\ifhat J_i, \ifhat N_{\ul jk}]=i \varepsilon_{ikl} \ifhat
N_{\ul jl}, \qquad [\ifhat \DSx_{\ul i}, \ifhat N_{\ul
jk}]=i \varepsilon_{ijl} \ifhat N_{\ul lk}, \nonumber \\
{} && [\ifhat N_{\ul ij}, \ifhat N_{\ul kl}] = -i
\left(\delta_{jl} \varepsilon_{\ul i \ul k \ul m} \ifhat\DSx_{\ul
m} + \delta_{\ul i \ul k} \varepsilon_{jlm}\ifhat
J_m\right),\nonumber
\\&& {}[\ifhat J_i, \ifhat D] = [\ifhat\DSx_{\ul
i}, \ifhat D] = [\ifhat N_{\ul ij}, \ifhat D] = 0,  \nonumber \\
&& [\ifhat J_i, \ifhat P_{\ul jk}]=i \varepsilon_{ikl} \ifhat
P_{\ul jl}, \qquad [\ifhat \DSx_{\ul i}, \ifhat P_{\ul jk}] =i
\label{bose commutators} \varepsilon_{\ul i \ul j \ul l} \ifhat
P_{\ul lk},
\\ {}&& [\ifhat N_{\ul ij}, \ifhat P_{\ul kl}]=i
\delta_{\ul i \ul k}\delta_{jl} \ifhat P_0 + i \varepsilon_{\ul
i\ul k\ul m}\varepsilon_{jln} \ifhat P_{\ul mn} ,  \nonumber \\
{}&& [\ifhat N_{\ul ij}, \ifhat P_0]=i \ifhat P_{\ul ij}, \qquad
[\ifhat D, \ifhat P_{\ul ij}]=i \ifhat P_{\ul ij}, \nonumber \\ &&
{} [\ifhat D, \ifhat P_0]=i \ifhat P_0, \qquad [\ifhat J_i, \ifhat
P_0] = [\ifhat \DSx_{\ul i}, \ifhat P_0]=0, \nonumber \\ &&
[\ifhat J_i, \ifhat K_{\ul jk}]=i \varepsilon_{ikl} \ifhat K_{\ul
jl}, \qquad [\ifhat \DSx_{\ul i}, \ifhat K_{\ul jk}] =i
\varepsilon_{\ul i \ul j \ul l} \ifhat K_{\ul lk}, \qquad \dots
\nonumber \\&& [\ifhat P_{\ul ij}, \ifhat K_{\ul kl}]= 2i \left( -
\delta_{\ul i\ul k}\delta_{jl} \ifhat D - \varepsilon_{\ul i\ul
k\ul m}\varepsilon_{jln} \ifhat N_{\ul mn} + \delta_{\ul i\ul
k}\varepsilon_{jlm}\ifhat J_m + \delta_{jl}\varepsilon_{\ul i\ul
k\ul m}\ifhat \DSx_{\ul m}\right), \nonumber \\ && {} [\ifhat
P_{\ul ij}, \ifhat K_0]=-2i \ifhat N_{\ul
ij}, \qquad [\ifhat P_0, \ifhat K_{\ul ij}]=-2i \ifhat N_{\ul ij}, \nonumber \\
{} && [\ifhat P_0, \ifhat K_0]= -2i \ifhat D.  \nonumber\eea

We will not investigate this construction, based on Heisenberg
operators, in more detail. We just note that realization of
conformal symmetry by quadratic Heisenberg operators has a number
of interesting features, e.g.\ operator $\DS{3}$ turns out to be
helicity operator (appearing here on the same footage with the
rest of conformal generators) generating both chirality symmetry
and electromagnetic duality symmetry \cite{ISduality}, while free
equations of motion show up as direct consequences of identities
connecting algebra generators \cite{ISequations}.\footnote{We call
attention that, had we utilized Heisenberg operators to directly
construct, in a standard fashion, an $so(2,4)$ conformal algebra
(instead of $sp(8)$) none of these would be present.}

In the next section we will replace starting Heisenberg algebra
with relations of parabose algebra, obtaining extension of
conformal supersymmetry. The case of Heisenberg algebra is
actually one representation (the simplest nontrivial) of that
superalgebra.

\section{\label{sec3}Parabose Operators and Extension of Conformal Supersymmetry}

Relations (\ref{conformal identification}, \ref{Jeleb
identification}) defining one basis of algebra $\quadalg$ can be
written the other way around in the following anticommutator
form: %
{ \refstepcounter{equation}
  \renewcommand{\arraystretch}{1.5}
  $$ \begin{array}{lr}
  \label{fermi anticommutators}
\{\hpp{\alpha},\hpp{\beta}\} = (\alpha_0)_{\alpha\beta}\, \ifhat
P_0 + (\alpha_{\ul ij})_{\alpha\beta}\, \ifhat P_{\ul
ij}, \\%
\{\hkk{\alpha},\hkk{\beta}\} = -(\alpha_0)^{\alpha\beta}\, \ifhat
K_0 + (\alpha_{\ul ij})^{\alpha\beta}\, \ifhat K_{\ul ij},
\qquad \qquad \qquad (\theequation) \\ %
\{\hkk{\alpha},\hpp{\beta}\} = (\alpha_0)^\alpha_{\ \beta}\,
\ifhat D + (\alpha_{\ul ij})^\alpha_{\ \beta}\, \ifhat N_{\ul ij}
+ (\sigma_{i})^\alpha_{\ \beta}\, \ifhat J_{i} + (\tau_{\ul
i})^\alpha_{\ \beta}\, \ifhat \DS{i}.
\end{array} $$
}%
In addition to the fact that operators $\hpp{}$ and $\hkk{}$
satisfy set of anticommutation relations, their commutators with
algebra $\quadalg$ generators: %
{ \refstepcounter{equation}
  \renewcommand{\arraystretch}{1.5}
  $$ \begin{array}{lr}
  \label{bose-fermi commutators}
{}[\ifhat J_i, \hpp{\alpha}] = -i (\frac{\ts
\sigma_i}{2})_\alpha^{\ \beta}\, \hpp{\beta},\qquad [\ifhat
\DS{i}, \hpp{\alpha}] = -i
(\frac{\ts \tau_{\ul i}}{2})_\alpha^{\ \beta}\, \hpp{\beta},\\%
{}[\ifhat D, \hpp{\alpha}] =  \frac{i}{2} \hpp{\alpha},\qquad
[\ifhat N_{\ul ij}, \hpp{\alpha}] = i (\frac{\ts \alpha_{\ul
ij}}{2})_\alpha^{\ \beta}\, \hpp{\beta},\qquad \qquad (\theequation)\\%
{}[\ifhat K_0, \hpp{\alpha}] = -i (\alpha_0)_{\alpha\beta} \,
\hkk{\beta}, \qquad [\ifhat K_{\ul ij}, \hpp{\alpha}] = i
(\alpha_{\ul ij})_{\alpha\beta} \, \hkk{\beta}, \\%
{} [\ifhat P_0, \hpp{\alpha}] = [\ifhat P_{\ul ij}, \hpp{\alpha}]
= 0
\mbox{\ \ \ (and similar $\hkk{}$ relations)} \\%
\end{array} $$
}%
show that $\hpp{\alpha}$ and $\hkk{\alpha}$ transform as Dirac
spinors under the Lorentz subgroup. To see this more clearly, we
can introduce the following (Majorana) representation of Dirac
matrices: \be \gamma_0 = i\tau_{\ul 2},\quad \gamma_i = \gamma_0
\;\! \alpha_{\ul 3i} = i\tau_{\ul 1}\sigma_i, \quad \gamma_5 = -i
\gamma_0 \gamma_1 \gamma_2 \gamma_3 = i \tau_{\ul 3},
\label{alpha and gamma connection}\ee %
so that Lorentz part of (\ref{bose-fermi commutators}) can be
written as $[\ifhat M_{\mu\nu}, \hpp{\alpha}] =
-i({\textstyle\frac 14} [\gamma_\mu, \gamma_\nu])_\alpha^{\ \beta}
\hpp{\beta}$ (where $\ifhat M_{ij}=\varepsilon_{ijk} \ifhat J_k$,
$\ifhat M_{i0} = \ifhat N_{\ul 3i} $).

Therefore, $\hpp{\alpha}$ and $\hkk{\alpha}$ are spinors whose
anticommutation relations close on algebra $\quadalg$. If there
were no commutation relations of starting Heisenberg algebra this
would have been a structure of a graded algebra. We now show that
it is possible to modify the whole idea as to obtain a true graded
algebra.

If the operators $\hkk{\alpha}$ and $\hpp{\alpha}$ satisfy
Heisenberg algebra, their linear combinations $\fto{\alpha} \bydef
\frac{1}{\sqrt 2} (\hkk{\alpha} + i \hpp{\alpha})$ and
$\ftoa{\alpha} \bydef \frac{1}{\sqrt 2} (\hkk{\alpha} - i
\hpp{\alpha})$, satisfy Bose algebra. However, in this section we
will change our starting point and assume that linear combinations
$\fto{\alpha}$ and $\ftoa{\alpha}$ satisfy parabose instead of
Bose algebra (i.e.\ that $\hkk{}$ and $\hpp{}$
satisfy some "para-Heisenberg" instead of Heisenberg algebra): %
\be [\{ \fto{\alpha}, \fto{\beta}\}, \fto{\gamma}] = 0, \quad [\{
\fto{\alpha}, \ftoa{\beta}\}, \fto{\gamma}] =
-2\delta^\gamma_\beta \fto{\alpha}.
\label{parabose algebra} \ee %
(Relations obtained from these by generalized Jacobi identities
and by hermitian conjugation are also implied.)

It is not difficult to verify that this change in starting algebra
does not influence relations (\ref{bose commutators}) and
(\ref{bose-fermi commutators}), if we keep the definitions
(\ref{fermi anticommutators}) [i.e.\ definitions (\ref{conformal
identification}, \ref{Jeleb identification})]. As the commutators
between $\hpp{\alpha}$ and $\hkk{\alpha}$ are no longer fixed, the
structure now presents a standard graded (mod 2) algebra.

Just like the algebra (\ref{bose commutators}) is extension of
conformal algebra, the graded algebra given by relations
(\ref{bose commutators}), (\ref{fermi anticommutators}) and
(\ref{bose-fermi commutators}) [isomorphic to $osp(1,8)$] can be
seen as an extension of $N=1$ conformal superalgebra. Indeed, by
setting $P_{\ul 1i} = P_{\ul 2i} = K_{\ul 1i} = K_{\ul 2i} =
N_{\ul 1i} = N_{\ul 2i} = \DS{1} = \DS{2} = 0$ we obtain standard
conformal superalgebra (comparison can be made using gamma
matrices (\ref{alpha and gamma connection}), identification
(\ref{conformal identification}) and by replacing: $\hpp{\alpha} =
Q_\alpha /\sqrt{2}$, $\hkk{\alpha} = \overline s^\alpha
/\sqrt{2}$, $\DS{3} = \frac 32 R$).

The connection in the opposite direction (from conformal
superalgebra to this extension) can be established if we notice
that anticommutator of two left-handed $\hpp{}$ operators
$\{\hpp{\eta}, \hpp{\xi}\}$, or of two right-handed operators
$\{\hpp{\dot \eta}, \hpp{\dot \xi}\}$ yields linear combination of
operators $P_{\ul 1i}$ and $P_{\ul 2i}$ (and similarly for
$\hkk{}$ operators). Therefore, our graded algebra can be seen as
a special conformal superalgebra where all anticommutators of
supersymmetry generators are allowed to be nonzero operators. (It
is enough to allow existence of nonzero $P_{\ul 1i}$ and $P_{\ul
2i}$ -- the introduction of the rest of the operators (\ref{Jeleb
identification}) is necessary to close the algebra.)

As we already announced in the introduction, by relaxing the
constraint $\{\hpp{\eta}, \hpp{\xi}\} = 0$ (i.e.\ $\{Q_{\eta},
Q_{\xi}\} = 0$) supermultiplets become infinite. Nevertheless, the
simple symmetry breaking assumption, discussed in the previous
section, breaks not only extra bosonic generators (\ref{Jeleb
identification}), but also the supersymmetry generators
$\hpp{\alpha}$ and $\hkk{\alpha}$. Since action of operators
$\hpp{\eta}$ and $\hpp{\dot \eta}$ change value of $\DS{3}$ for
$\frac 12$, each following member of a supermultiplet would gain
higher and higher mass, whereas the low-energy space-time symmetry
would be given by Poincare group.


\section{\label{sec4}Conclusion}

In this paper we analyzed an extension of the conformal
supersymmetry. It is interesting that, although the proposed
symmetry is higher and mathematical structure thus richer, the
algebra relations are simplified. Namely, commutators of bosonic
with fermionic operators (\ref{bose-fermi commutators}) are
nothing more than simple relations of parabose algebra written in
a complicated basis. And the fermionic anticommutators (\ref{fermi
anticommutators}) are relations that describe this new basis, so
these relations can be seen as a specific naming convention for
linear combinations of $\hpp{}$ and $\hkk{}$ anticommutators. The
idea is that this complicated basis becomes physically relevant
due to the symmetry breaking, analyzed in section \ref{sec2}. The
relatively simple symmetry breaking is therefore responsible not
only for reduction of the starting symmetry and for introduction
of mass splitting, but also for superficial complexity that hides
simplicity of the starting parabose algebra. Bosonic algebra
$\quadalg$ relations (\ref{bose commutators}) are direct
consequence of (\ref{fermi anticommutators}) and (\ref{bose-fermi
commutators}).

From the perspective of this higher symmetry, those relations of
standard conformal superalgebra that set some of the
anticommutators to zero appear as a kind of artificial constraints
-- constraints that are, in this picture, consequences of a
symmetry breaking. This fact that some linear combinations of
anticommutators are zero makes it impossible to see
anticommutators of fermionic generators simply as a naming
convention, as it was possible for (\ref{fermi anticommutators}).
On the other hand, extension from the conformal superalgebra to
the symmetry discussed here can be done by allowing all
anticommutators of supersymmetry generators to be nonzero
operators. By doing so we end up with an algebra determined by
only two parabose relations (\ref{parabose algebra}).

\end{document}